\documentclass[pre,aps,epsf,twocolumn]{revtex4}
\usepackage{psfig}
\newcommand{\be}{\begin{equation}}
\newcommand{\ee}{\end{equation}}

\begin{document}

\title{Nonlinear response and discrete breather excitation
in driven micro-mechanical cantilever arrays.}
\author{ P. Maniadis and S. Flach}
\address{ Max-Planck-Institut f\"ur Physik Komplexer Systeme\\
N\"othnitzer Str. 38, D-01187 Dresden, Germany.}
\date{\today}

\begin{abstract}
We explain the origin of the generation of discrete breathers (DBs)
in experiments on damped and driven micromechanical cantilever arrays
(M.Sato et al. Phys. Rev. Lett. {\bf 90}, 044102, 2003).
Using the concept of the nonlinear response manifold (NLRM) we provide
a systematic way to find the optimal parameter regime in damped and driven 
lattices where 
DBs exist. Our results show that  DBs appear via a new instability of the NLRM
different from the anticipated modulational instability (MI) known
for conservative systems. 
We present several ways of exciting DBs, and compare 
also to experimental studies of exciting and destroying DBs
in antiferromagnetic layered systems.
\end{abstract}


\maketitle

The existence and the properties of intrinsic localised modes (ILMs) or
discrete breathers (DBs) in nonlinear 
lattices have been investigated thoroughly during the last years
(see \cite{DB_rev} and references therein).
In addition an impressive number of various  
experiments during the last years have verified the existence of these modes
in many systems like micro-mechanical cantilever arrays \cite{cant1,cant2},
antiferromagnets \cite{NatureSievers,magnet}, 
Josephson junction arrays \cite{Trias,Ustinov},
coupled optical waveguides \cite{wavegides},
atomic vibrations of highly nonlinear
materials \cite{PtCl,Ham1,Ham2} and Bose-Einstein condensates 
on optical lattices \cite{BEC}.

The bulk of the central theoretical results has been achieved for
conservative systems. One good reason for that is the complexity of
the DB properties. While DBs typically persist under the influence
of weak dissipation (which should also include an energy pumping mechanism),
various realisations of dissipation 
(dc driving, ac driving, fluctuations,
linear versus nonlinear damping etc) modify DB properties in a specific way,
turning the limiting conservative case into an ideal starting playground
for setting a coherent frame of the understanding of their properties. 
Since most of the experimental studies face dissipation, each case may
call for a specific additional theoretical study.  

DB observation in antiferromagnets \cite{NatureSievers,magnet} and
cantilever arrays \cite{cant1,cant2} involved the excitation of the system
with spatially homogeneous external fields, triggering a
spatially inhomogeneous  system state 
via some inherent instability. For conservative systems the 
modulational instability (MI) of band edge plane waves is known 
to provide such a path.
Especially for the case of driven cantilevers, the MI
approach for conservative systems was used to design an experimental system 
of alternating short and long cantilevers \cite{cant1,cant2}. 
The results we present below show that DBs appear
via a completely different instability. The MI can be used for
energy pumping and consequent overcoming of dynamical barriers.
However several other routes with a similar outcome can be exploited.
We demonstrate that by exciting DBs in arrays with identical cantilever length. 

In order to study the driven cantilever system we use the model equations
in \cite{cant1,cant2} and 
introduce the dimensionless time $t\rightarrow t/t_0$ and
displacement $x_l\rightarrow x_l/a_0$ where $t_0$ and $a_0$ 
correspond to a characteristic time and length of the system.
The equations of motion describing the cantilever system can be then written as 
a system of coupled anharmonic damped and ac driven oscillators
\be
\ddot{x}_l+\gamma \dot{x}_l+a_2x_l +a_4x_l ^3-C(x_{l+1}+x_{l-1}-2x_l)=A(t)
\label{eqmot1}
\ee
The oscillator displacements $x_l$ describe 
the deflection angle of the $l$-th cantilever
from its equilibrium position. The hard-type anharmonicity tends to increase
the oscillation frequencies with growing amplitudes.
This model neglects the influence of longer than  nearest neighbour 
interaction range,
which is not crucial for the understanding of the main qualitative 
DB properties.
The dimensionless parameters are related with  
the ones of the experiment in ref.\cite{cant1,cant2}: 
$\gamma =t_0/\tau$,
$a_2=k_2t_0^2/m$,
$a_4=k_4t_0^2a_0^2/m$,
$C=k_lt_0^2/m$ 
and $A=\alpha t_0^2/a_0$.
Using the experimental values in \cite{cant1,cant2} and  setting $a_2=a_4=1$, 
we find $t_0=1.34238\cdot 10^{-6}$sec and
$a_0=2.46\cdot 10^{-5}$m.
The friction and coupling parameters become $\gamma =1.534\cdot 10^{-4}$
and $C=0.07953$. The spatially uniform ac driving $A(t)$ in (\ref{eqmot1})
is generated by a corresponding piezoelectric crystal vibration in the 
original experiments.

Neglecting the damping $\gamma$, the ac driving $A(t)$ and the nonlinear
force terms in (\ref{eqmot1}) one readily derives the only possible
solutions, namely plane waves $x_l \sim {\rm e}^{i(\omega_q t - ql)}$
with the linear dispersion relation 
$\omega _q =\sqrt{1+4C\sin^2(q/2)}$ relating the plane wave frequency $\omega_q$
to its corresponding wave number $q$. Reinstating the nonlinear force terms
in (\ref{eqmot1}) leads to two conclusions \cite{DB_rev}: 
i) discrete breathers, 
i.e. time-periodic and spatially localised solutions 
$x_l(t)=x_l(t+T_b)\;,\;x_{|l| \rightarrow \infty} \rightarrow 0$
exist for frequencies 
$\Omega_b=\frac{2\pi}{T_b} > \omega_{q=\pi}$; ii) the $q=\pi$ plane
wave mode turns (modulationally) unstable at amplitudes 
$\sim 1/N$ where $N$ is the number
of cantilevers, and DBs bifurcate from this plane wave mode along this very 
instability route.
What can we expect if both damping and ac driving are added as well?  
DBs persist without much change, thus
the ac driving frequency choice $\omega_d > \omega_{q=\pi}$ is well reasoned.
However instabilities of conservative systems may turn sensitive to effects
of damping. Assuming that the MI of the staggered $q=\pi$ mode is the track 
to follow,
the choice of a nonstaggered ac driving is not appropriate. 
Note that the experimental
design of alternating short and long cantilevers splits the spectrum
$\omega_q$ into two bands separated by a gap. In addition it formally 
transforms the $q=\pi$ mode of the system with identical cantilevers 
into the $q=0$ mode of the  
upper band for the system with alternating cantilever length. 
Yet the staggered character
of this mode is of course preserved inside each unit cell which contains 
now two
neighbouring cantilevers. Thus the mismatch between the staggered 
MI mode and the
nonstaggered ac driving remains also for systems
with alternating cantilever length. 

With the above parameters $1 \leq \omega_q \leq 1.1481$. The narrowness
of the band as compared to the characteristic frequency values is due
to the weak coupling constant $C$. Let us start then from the 
uncoupled limit. In that case each oscillator evolves independently.
Assuming for the moment small amplitudes of all oscillators, the presence
of a weak driving will (after some transient time) bring them all
into a unique oscillatory state, thus all oscillators will move in phase
(with each other). Consequently we have to study the stability of
a {\it nonstaggered} extended time-periodic state for weak coupling as well,
no matter whether the frequency of the driving is located above or below
the band $\omega_q$.

For a periodic driving of the form $A(t)=A_0\cos(\omega _d t)$, 
the equation (\ref{eqmot1})
support periodic solutions.
It is easier to study first the properties of these solutions at zero friction
($\gamma =0$), and then examine the modifications when $\gamma \neq 0$.
Newton method \cite{DB_rev,ma} was used for the 
tracking of periodic solutions 
with frequency equal to the external driving of the form: 
$x_l(t)=x_{0,l}\cdot f(\omega _d t)$ where $x_{0,l}$ is the amplitude of the 
oscillation, and $f(\omega _d t)$ is a periodic function with period $2\pi$.
The Newton scheme was initiated with a very small driving amplitude 
($A_0 \simeq 0$),
and the response of the system (i.e. the amplitude of the oscillations
$x_{0,l}$) was followed as $A_0 $ was varying. Thus we reconstruct the full
Nonlinear Response Manifold (NLRM) \cite{Kopidakis2}.

The NLRM close to the origin follows from linearising
the equation of motion (\ref{eqmot1}). The exact solution  
of the system in this limit is $x_l(t)=x_{0,l}\cos(\omega _d t)$ with
$x_{0,l}=-A_0/(\omega _d ^2-1)$.
This is a homogeneous branch (HB) 
(all the oscillators are in phase). There is a 
phase difference of $\phi =\pi$ between the driving and the 
response of the system.
The NLRM is symmetric around the origin due to  
$x_l(t+\pi/\omega_d)=-x_l(t)$ and $A(t+\pi/\omega_d)=-A(t)$.

\begin{figure}[h]
\centerline{\hbox{\psfig{figure=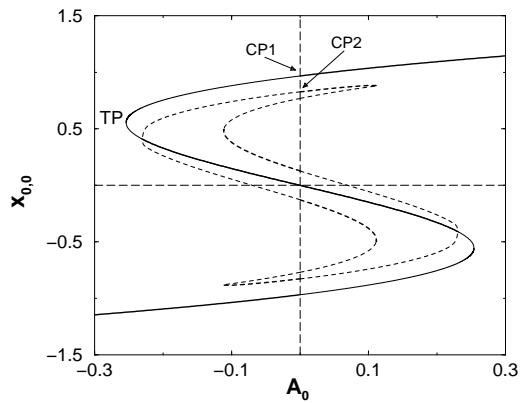,width=6.8cm,angle=270}}}
\caption{The nonlinear response manifold (NLRM) for $C=0.0079$ and 
$\omega _d=1.3$. Solid line - homogeneous branch, 
dashed line - various breather branches for $x_{0,0}$. 
Thick lines correspond to stable parts of the manifold, thin -
to unstable ones.}
\label{fig1}
\end{figure}

Increasing $|A_0|$, 
the displacement $x_{0,0}$ of HB increase,
up to the first turning point (TP)
(Fig.\ref{fig1}).
After the turning point $x_{0,0}$ continue to increase, while the
driving amplitude decreases down to zero at the 
first crossing point (CP1). This crossing point corresponds to 
a homogenous solution with all the 
oscillators are oscillating with the same amplitude $x_{0,0}$ and zero driving.
This state exists due to the frequency increasing, hard anharmonicity term
in the equations of motion and because $\omega_d > \omega_q$.
The NLRM manifold of the HB continues further from CP1 with again nonzero $A_0$,
but the solution is now in phase with the driving. 

The Floquet stability analysis \cite{DB_rev} of the HB
reveals that close to the origin the manifold is linearly stable. 
An instability appear before the TP and the HB  becomes unstable.
The NLRM of the homogeneous branch continues to be unstable between the 
TP and the CP1.
After the CP1, the homogeneous branch turns stable again (see Fig.\ref{fig1}).

\begin{figure}[h]
\centerline{\hbox{\psfig{figure=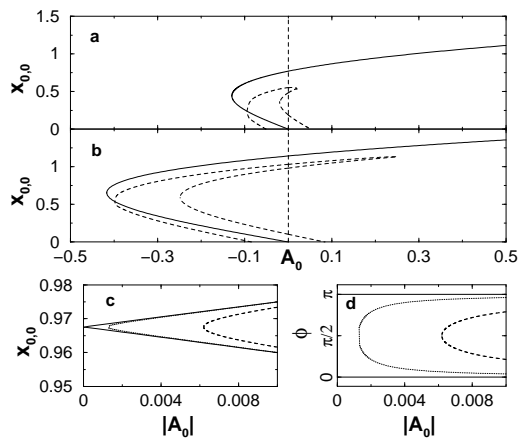,width=6.8cm,angle=270}}}
\caption{(a,b) The NLRM for different values of the driving frequency:
(a) $\omega _d=1.2$,
(b) $\omega _d=1.4$.
The parameters are the same used in Fig.\ref{fig1}.
(c,d) the response $x_{0,0}$ and the phase difference $\phi$ of the manifold
in the neighbourhood of the first crossing point for different values of the 
friction. Continuous line corresponds to $\gamma =0$, dotted line to 
$\gamma =0.001$ and dashed line to $\gamma = 0.005$.}
\label{fig2}
\end{figure}

The instabilities of the HB mark the 
bifurcation of spatially inhomogeneous solutions - various breather states -
off the HB. 
For a system of $N$ oscillators, at the first instability, 
$N$ new branches of the NLRM appear. Each of these branches corresponds to 
a single breather centred at a different site on the lattice. 
In Fig.\ref{fig1} we show two of these branches
with dashed lines. 
Starting from the bifurcation, the 
displacement of the central oscillator increases, while $|A_0|$ decreases.
For $A_0=0$ the manifold 
passes through the second crossing point (CP2). This branch of the manifold 
corresponds to a breather, and is stable for $A_0>0$ and 
unstable for $A_0<0$.
After that, the manifold turns, and passes through further crossing points.
Thus for fixed $\omega_d$ we have multistability for
small enough $A_0$ with a large amplitude and small amplitude HB coexisting
together with breather states. For large enough $A_0$ only the large amplitude
HB survives.

The NLRM depends on the driving frequency $\omega _d$.
In Fig.\ref{fig2} we show a part of the NLRM for two 
different values of the driving frequency. In Fig.\ref{fig2}(a) 
$\omega _d=1.2$ and in Fig.\ref{fig2}(b) $\omega _d=1.4$.
With increasing frequency 
the slope of the manifold decreases, the crossing points appear for larger
values of $x_{0,0}$, as a result the bifurcations 
and the turning points appear for larger values of $|A_0|$.

The properties of the manifold are slightly modified 
due to the presence of friction ($\gamma \neq 0$).
\begin{figure}[h]
\centerline{\hbox{\psfig{figure=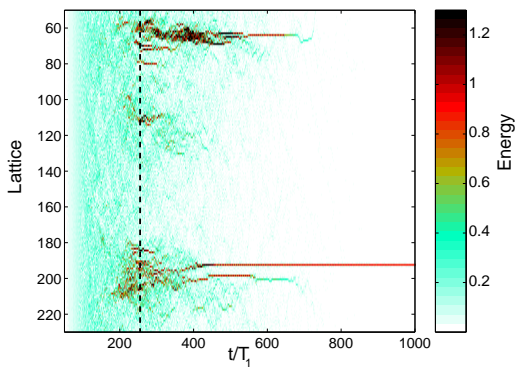,width=6.8cm,angle=0}}}
\caption{ Energy as a function of time. The driving frequency is initially 
ramped from $\omega _d=1.14$ to $\omega _d=1.4$, and then it remains fixed
(marked by the vertical dashed line).
Energy pumping is observed during the ramping. At the end of the ramping 
process, breathers are 
created, and one of them locks to the driving and survives.  
$T_1 =20\pi /\omega _d$.}
\label{fig3}
\end{figure}
The NLRM between the origin and the crossing point (CP1) for $\gamma=0$ 
is in anti-phase
with the driving, while after the crossing point the NLRM is in phase with 
the driving. For $\gamma=0$ therefore there is a phase difference
of $\phi = \pi$ in the response of the system around a CP.
When a small friction is introduced in the system, it creates 
an extra small phase difference between the driver and the 
response. This phase difference between the driving and the response 
increases in the neighbourhood of the CP. 
The result is that the solution at the CP
disappears due to friction, but there is a smooth transition 
between the two different branches of the manifold. 
The phase difference $\phi$ varies smoothly
from $\pi - \epsilon _-$ to $0+\epsilon _+$ where $\epsilon _\pm$
correspond to the small phase differences created by the friction on the large 
and low amplitude branch respectively, far from
the CP. For weak damping this transition occurs very close to the 
CP. 
The smooth transition between the two branches of the manifold in the 
neighbourhood of (CP1) as a function of $|A_0|$
is shown in Fig.\ref{fig2}(c) for different values of the friction. 
The smooth transition in the phase difference $\phi$ is shown 
in Fig.\ref{fig2}(d).
\begin{figure}[h]
\centerline{\hbox{\psfig{figure=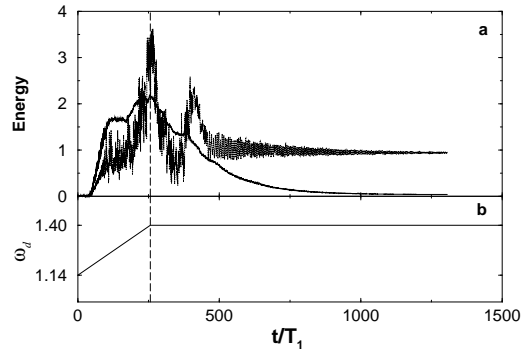,width=6.8cm,angle=270}}}
\caption{(a) Average energy per oscillator as a function of time 
(solid line, scaled by a factor of 10 for better observation) 
and the energy of the locked oscillator (strongly fluctuating dotted line).
The parameters and the frequency ramping are like in Fig.\ref{fig3}.
(b) The frequency ramping scheme. After the vertical dashed line the driving 
frequency remains constant.}
\label{fig4}
\end{figure}
Far from the CP, the properties of the manifold 
are only slightly modified by the nonzero friction.

\begin{figure}[h]
\centerline{\hbox{\psfig{figure=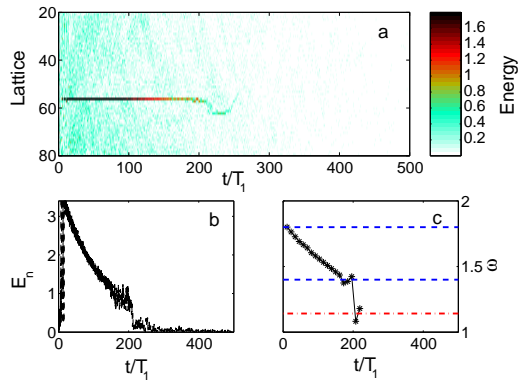,width=6.8cm,angle=0}}}
\caption{(a) Energy density evolution
for a sequence of different driving (see text for details);
(b) Time dependence of the DB energy observed in (a);
(c) Time dependence of the DB frequency observed in (a).
The two dashed lines mark the frequencies $\omega _1$ and $\omega _2$.
The dashed-dotted line corresponds to band edge frequency 
$\omega_{\pi}$.
}
\label{fig5}
\end{figure}

As follows from Fig.\ref{fig2}, the stable breather branch is
not connected to the stable HB of the NLRM. Thus we can exclude
an easy way of exciting the system in the HB and tuning some
parameter (amplitude or frequency of the driving) such that
we continuously join the breather branch. This calls for
stochastic excitations of the system to enforce a nonzero probability
to end up in a breather state when initially starting from a HB.
Similar to the experiments \cite{cant1,cant2} we performed a ramping 
of the
frequency $\omega_d$ from 1.14 (inside the band) to 1.4 (above the band)
for fixed amplitude $A_0=0.008$ (cf. Fig.\ref{fig4}(b)). 
According to Fig.\ref{fig2}
the final state corresponds to the multistability
domain of the NLRM. If the system was initially at rest,
the final state is observed to be the low amplitude HB.
However, adding some nonzero noise,
uniformly distributed in the interval $(-0.05,0.05)$,
in the initial conditions of the cantilever
deflection angles, the ramping process excites 
a unstable plane wave, which provides a finite time energy pumping
due to MI (Fig.\ref{fig3}). The MI acts as a noise amplifier,
which allows for the generation of hot spots or breather
precursors. After the ramping finishes, most of these hot spots
decay back into the low amplitude HB, while some
of them may lock to the driver and turn into a stable breather.
The average energy per oscillator is shown in Fig.\ref{fig4}(a)
as a function of time. 
During the ramping, there is a frequency window, where energy is 
pumped into the system, and then the system relaxes due to dissipation. 
The dotted line shows the energy of the locked breather site.

In order to verify that the energy pumping mediated by the MI in the
experiments is only needed to amplify the noise, we 
perform simulations in the absence of a frequency ramping,
but with the initial fluctuations being larger 
(Gaussian distribution of initial displacements with zero mean and
variance $D=0.5$).
Keeping the frequency and the driving amplitude fixed
($\omega _d=1.4$ and $A_0=0.008$) we observe the creation and 
locking of breathers similar to Fig.\ref{fig5}. 
Similar simulations with alternating long and short cantilevers, 
reveal the same behaviour. 

The NLRM study shows, that breathers can be obtained in a driven and
damped system by carefully choosing the frequency (outside the phonon
band) and amplitude of
the driving force (not being too large) 
such that the NLRM is activated inside the multistability
domain. We have shown that localised breathers emerge from the
homogeneous solution via an instability
of the NLRM completely different from the expected MI picture.
The fact that the stable breather branch is disconnected from
the stable HB, implies that fluctuations have to be used
in order to perform a crossover from one to the other.
There exist various pathways of generating breathers,
either by frequency ramping and exploiting the MI, 
or by initially strongly exciting the system.
These pathways have to be designed in such a way that 
the system will tend to the small amplitude HB of the NLRM,
with some fluctuations growing into hot spots which finally
transform into breather states. 

The NLRM can be also used to study the breather excitation
in antiferromagnetic systems \cite{NatureSievers,magnet}
where external driving was used as well. The observed breather destruction
in these experiments is similar to the process
presented in Fig.\ref{fig5}. 
In analogy with the experiment a strong driving at frequency $\omega _1=1.8$
with $A_1=0.8$ and duration of $t=20T_1$
is used, together with initial displacements being uniformly
distributed in the interval $(-0.25,0.25)$. 
As a result a DB is formed (Fig\ref{fig5}(a)). 
A subsequent long driving
at a frequency $\omega _2=1.4$ (being located closer to the spectrum
$\omega_q$) is exciting the system with a small amplitude
$A_2=0.008$. Because of the frequency mismatch
the breather with $\omega _1$ will start to decay, but tend to 
slow down the decay when its frequency is close to $\omega _2$
(Fig\ref{fig5}(c)). Similar to the experimental case we
add a third weak driving signal with $\omega _3=2$ and amplitude $A_3=0.0004$.
That hinders the locking of the DB to $\omega _2$. 
Right after the DB frequency passes $\omega _2$, its relaxation speeds up
and the excitation is destroyed very quickly (Fig.\ref{fig5}(b),
cf. also Fig.3 in \cite{NatureSievers}. The observed energy release
is fixed by the energy value of a DB locked to the $\omega _2$ driving
and explains the experimental observation of equal height steps
in the relaxation of DBs. At the same time we interprete the observed
smooth decrease of the emission signal in Fig.3 in \cite{NatureSievers}
as a process of slow relaxation of DBs with frequency $\omega _1$
towards DBs with frequency $\omega _2$.
\\
\\
Acknowledgements. We thank A. J. Sievers and U. Schwartz for
helpful discussions.

\end{document}